\documentstyle[emulateapj]{article}
\begin{document}
\title{\bf The Power Spectrum of Rich Clusters on Near-Gigaparsec Scales}

\author{Christopher J. Miller and David J. Batuski}
\affil{Department of Physics \& Astronomy, University of Maine}

\begin{abstract} 
Recently, there have been numerous analyses of the redshift space power spectrum
of rich clusters of galaxies. Some of these analyses indicate a \lq\lq bump\rq\rq~
in the Abell/ACO cluster power spectrum around $k = 0.05h$Mpc$^{-1}$.
Such a
feature in the power spectrum excludes most standard formation models and indicates
possible periodicity in the distribution of large-scale structure. 
However,
the data used in detecting this peak include clusters with estimated redshifts
and/or clusters outside of Abell's (1958) statistical sample, {\it i.e.} $R = 0$ clusters. 
Here, we present estimates of the redshift-space power spectrum for a newly expanded
sample of 637 $R \ge 1$ 
Abell/ACO clusters which has a constant number density to z = 0.10 in the Southern Hemisphere
and a nearly constant number density to z = 0.14 in the Northern Hemisphere. 
The volume sampled, $\sim 10^8h^{-3}$Mpc$^{3}$, is large
enough to accurately calculate the power per mode to scales approaching $10^3h^{-1}$Mpc.
We find the shape of the power spectrum is a power-law on scales $0.02 \le k \le 0.10h$Mpc$^{-1}$,
with
enhanced power over less rare clusters such as APM clusters. The power-law here follows $n = -1.4$.
The power spectrum is essentially featureless,  
although we do see a dip near $k = 0.04h$Mpc$^{-1}$ which cannot be considered
statistically significant based on this data alone.
We do not detect a narrow peak at $k \sim 0.05h$Mpc$^{-1}$ and
there is no evidence for a turn-over in the power spectrum as has been previously reported.     
We compare the
shape of the Abell/ACO rich cluster power spectrum to various linear models.

\end{abstract}

\section{Introduction}
There has recently been a renewed interest in accurately determining the
power spectrum of matter distribution scales greater than 100$h^{-1}$Mpc;
in part due to the increased
number of clusters with measured redshifts and the large volumes they trace.
The power spectrum for
the galaxy distribution has been determined many times for many different classes
of galaxies. However, most galaxy surveys lack the volume necessary for the
accurate quantification of power on large-scales ({\it e.g.} the Las Campanas Redshift Survey 
(Lin {\it et al.} 1996-hereafter LCRS) or the Stromlo-APM survey (Tadros \& Efstathiou 1996).
A summary of the results from these analyses is that the redshift-space
power  spectra roughly agree on scales $\lambda$ ($ = 2\pi/k$) $< 100 h^{-1}$Mpc.
In this region, $P(k) \propto k^n$ and $n \sim -2$. (Of course, the amplitude of the
power spectra depends on the samples of galaxy examined which provides strong evidence for
a luminosity bias (see e.g. Vogeley et al. 1992 and Park et al. 1994)).
However, on scales $\lambda > 100h^{-1}$Mpc, there is much less
agreement. For example, some galaxy samples, such as from the  LCRS
and the Automated Plate Machine (APM) 2d and 3d surveys 
show a  broad flattening
around $k = 0.05 h$Mpc$^{-1}$ although no distinct maximum can be found within
convincing statistical bounds (LCRS;
Tadros and Efstathiou 1996; Peacock 1997; Gatzanaga \& Baugh 1998).
However, Landy {\it et al.} (1996) find a distinct peak in $P(k)$ for a 2 dimensional
analysis of the LCRS and Broadhurst {\it et al.} (1990)
find a peak near $\lambda = 130h^{-1}$Mpc in
a deep pencil beam survey.

Some Abell/ACO cluster analyses 
also show a peak around
$ k \sim 0.05 h$Mpc$^{-1}$ (Retzlaff {\it et al.} 1998; Einasto {\it et al.} 1997-hereafter E97).
Yet other cluster analyses only show a smooth turnover
in the power spectrum to its scale-invariant ($n = 1$) form. For instance,
the APM clusters, examined by Tadros, Efstathiou and Dalton (1998-hereafter TED98)
show a maximum in $P(k)$ at the smaller value of
$k \sim 0.03h$Mpc$^{-1}$ and no distinct \lq\lq bump\rq\rq~ at $k = 0.05h$Mpc$^{-1}$.
Also, Peacock \& West (1992) and Jing \& Valdarnini (1993) find a break in the Abell cluster
power spectrum near $k = 0.05h$Mpc$^{-1}$, but no distinct peak in power.
An excellent review of the power spectra for different galaxy species can be found in 
Einasto {\it et al.} (1999-hereafter E99). E99 determine a mean
power spectrum for all galaxies for a large range in wavenumber. They do this by 
using the APM 2d power spectra on small scales, and by 
averaging over numerous samples on large scales and then normalizing to the APM 2d power. 

If a narrow peak in power near $k \sim 0.05h$Mpc$^{-1}$ is a real feature of the power spectrum
in general, most current models of structure formation (in the quasi-linear regime) become invalid
(E99). While baryonic signatures can produce features in the power spectrum,
those features are oscillatory and they do not produce a singular, narrow peak as seen
in some of the current data. Eisenstein et al. (1998) examined this prospect and found
that no selection of cosmological parameters reproduces the power spectrum as seen in E97.
However, Gramann \& Suhhonenko (1999) suggest that an inflationary scenario with a scalar field
having a localized step-like feature can reproduce the power spectrum of clusters.
However, in this work, we show that the peak in the cluster power spectrum is not
present in larger (in volume and in number)
cluster samples after excluding less reliable data (such as $R=0$ Abell/ACO clusters
and clusters with estimated redshifts).

Our aim in this paper is to provide an estimate for the power spectrum of Abell/ACO
clusters that is based on a complete and fair sample. Both Retzlaff {\it et al.} and
E97 use $R =0$ clusters in their determination of $P(k)$. 
Einasto {\it et al.} (1994) have argued that $R =0$ clusters
do not contaminate studies of large-scale structure because the multiplicity of
superclusters is independent of richness and the mean separation distances for
$R =0$ and $R \ge 1$ clusters are very similar.
However,  $R = 0$ clusters were not cataloged in a systematic way and were never 
meant to be examined in a statistical manner due to their incompleteness (Abell 1958).
In addition, many researchers have found line-of-sight anisotropies in $R = 0$ cluster
samples (Sutherland 1988; Efstathiou {\it et al.} 1992; Peacock \& West 1992).
Therefore, the use of $R =0$ clusters in the determination of $P(k)$
is highly suspect. E97 have also used a large
number (435 out of 1305 clusters) of estimated redshifts in their
determination of $P(k)$. We also suspect that
E97 used a large number of cluster redshifts with only one measured galaxy. Miller {\it et al.}
(1999a) show that cluster velocities with one measured galaxy are in error by more than $2500$ km s$^{-1}$
14\% of the time.
Of course, estimated redshifts are only
accurate to at best 25\%. Thus, the statistical certainty of any large-scale structure
analyses based on the cluster samples with a large number of estimated or poorly determined
redshifts must also be taken with caution.

\section{The Cluster Sample}
We examine Abell/ACO clusters across the entire sky excluding the galactic plane 
{\it i.e.} $|b| > 30^{\circ}$. We only consider $R \ge 1$ clusters (with
measured redshifts) since they
were defined by Abell (1958) as members of his statistically complete sample.
Recently, Miller {\it et al.} (1999a,b) examined similar subsets of $R \ge 1$ clusters
for projection effects, line-of-sight anisotropies, and spatial correlations. We
summarize their results below.

The Abell/ACO $R \ge 1$ cluster dataset has significant advantages over other cluster
samples (including those with $R =0$ clusters as well as APM clusters). With the
advent of multi-fiber spectroscopy, nearly all rich Abell/ACO clusters within $z=0.10$ now have
multiple galaxy determined redshifts (Slinglend {\it et al.} 1998; Katgert {\it et al.}
1996). Multiple redshifts have allowed for more accurate determinations of the extent
of projection effects and Miller {\it et al.} 1999a report that at most, 10\% of
Abell/ACO clusters suffer from moderate to severe foreground/background contamination.
The lack of projection effects
for $R \ge 1$ clusters is also apparent from the 89\% X-ray emission
detection rate by Voges {\it et al.} 1999.
Miller {\it et al.} (1999a,b) also show that there is very little line-of-sight anisotropy
in the $R \ge 1$ Abell/ACO cluster samples - comparable to the APM cluster
catalog (Dalton {\it et al.} 1994)
This is in sharp contrast to $R \ge 0$
samples and even some modern X-ray selected/confirmed cluster samples (see e.g. Efstathiou
et al. 1992, Peacock \& West 1992, and Miller et al. 1999b).

Vogeley (1998) recently pointed out how Galactic extinction could
add ``false'' power to structure analyses based on large galaxy
samples (such as the Sloan Digital Sky Survey). While clusters should
not affected as strongly as individual galaxies, it is still worth examining
extinction effects within our cluster sample.
In 1996, Nichol and Connolly used the Stark {\it et al.} 1992 HI maps to report
that some samples of Abell clusters significantly anti-correlate with
regions of high galactic neutral hydrogen density. 
Recently, Schlegel, Finkbeiner, and Davis (1998) have
created HI extinction maps of the entire sky with much greater resolution than the Stark HI
maps. We use these new maps to re-examine
and confirm the Nichol and Connolly results. We also examine a volume-limited ($z=0.10$) sample
of Abell/ACO clusters.
Using a Kolmogorov-Smirnov (K-S) test, we compare the E(B-V) extinctions for positions
centered on the Abell/ACO clusters to E(B-V) extinctions for several thousand randomly
selection positions. We find that the probability that our clusters were drawn
from a random selection of E(B-V) extinctions is 10\%.  
In other words, the average extinction
within our Abell/ACO clusters is smaller than for the random positions, but
not significantly so. For comparison, Nichol and Connolly found only a 2\% probability that the
Postman, Huchra, and Geller (1992) Abell/ACO clusters (with $|b| \ge 30^o$ and $R \ge 1$) were
drawn from a random sampling of E(B-V) extinctions. 
The effect that galactic extinction would have on a power spectrum should not be 
as strong for clusters as it would be for galaxies. Cluster galaxies have
a wide range of magnitudes, and while some dimmer galaxies within a cluster may be missed due
to extinction, the majority of bright galaxies will still be counted. When we created
our volume-limited samples, we are including those clusters that appear dim as a result of
galactic extinction (as opposed to a magnitude-limited survey which would exclude those
clusters). The lack of statistically significant evidence that our clusters
are corrupted by extinction, and the use of a volume-limited sample (with $|b| \ge 30^o$),
convinces us that
we can ignore any extinction effects in our analyses.
However, to be certain that extinction is not altering the shape or amplitude of
our redshift-space power spectrum, we will model the extinction distribution
of our clusters in our random catalogs for one of our two PS estimation methods (see method (b) below).

An additional argument for the completeness of $R \ge 1$ Abell/ACO clusters is
provided by their spatial number density as shown in Figure 1. 
We use clusters of all magnitudes and use the same methods 
as Miller {\it et al.} (1999a) to calculate and
bin the cluster number densities. 
Notice in Figure 1 that the sample
has a nearly constant density out to $z = 0.10$
and that the density
only drops by a factor of $0.58$ out to $z = 0.14$.  
[Note: The bump in the density at $z \sim 0.07$ is mostly due to
the Corona Borealis Supercluster.]
In Figure 1, we fit three different functions to the number density:
a three-parameter number function (as in FKP), a power-law for $z \ge 0.10$,
and a step-function.  The best-fit produces $\chi^2_{red} = 2$  for the number function. 

Using cluster redshifts from the literature as well as $\sim 100$ as yet unpublished
redshifts from the MX Survey Extension (Miller {\it et al.} 2000),
we have created a sample of 637 $R \ge 1$ Abell/ACO clusters with $|b| \ge 30^{\circ}$.
The MX Survey provides a much deeper (in both magnitude and in redshift) catalog
of Northern Hemisphere cluster redshifts than is currently available for the Southern Hemisphere ACO clusters
(see e.g. Miller et al. 1999; Katgert et al. 1996).
Therefore, we exclude any cluster
beyond $z = 0.10$ in the south ($\delta \le -27^{\circ}$) and beyond $z = 0.14$ in the
north ($\delta \ge -27^{\circ}$).
Several researchers have noted discrepancies between the richness counts of
the Abell and ACO catalogs (see Miller et al. (1999) for a discussion).
Therefore, we also measure $P(k)$ for a subset of our data that
excludes all ACO clusters with
$N_{gal} < 55 $ (where $N_{gal}$ is the number of galaxies used to determine the richness as
given in ACO and $N_{gal} \ge 50$ corresponds to $R \ge 1$). This richness cut excludes
30 ACO clusters from our sample.

This is the largest cluster sample compiled to date for large-scale structure analyses.
The survey volume covers $1.2\times10^8h^{-3}$Mpc$^3$ and is nearly four times larger than
the APM cluster survey  (Dalton {\it et al.} 1994)
and the Retzlaff {\it et al.} (1998) Abell/ACO survey.
Additionally, only  $\sim$ 10\% of our cluster redshifts are
based on one measured galaxy redshift.
We calculate distances to the clusters using
a Friedmann Universe with $q_0 = 0$ and $H_0 = 100$ km s$^{-1}$ Mpc$^{-1}$. 
The choice of $q_o$ makes little difference in
our results (see also Retzlaff {\it et al.} 1998).

When a cluster dataset goes as deep as the one used here, and has been
created in a somewhat piecemeal fashion, we must be very 
confident that the cluster observations used in this sample are more or less isotropic in volume, and that
we are not including large sections of the sky that go deeper (in magnitude) than others.
We address this concern 
figuratively in Figure 2, by examining the fraction of observed to total cataloged clusters.
In this sky plot (in galactic coordinates), we
show Abell clusters with $z=0.10$ (filled circles), Abell clusters within
$0.10 < z \le 0.14$ (open circles), and ACO clusters within $z=0.10$ (stars).
We can divide the sky into quadrants with two sections in the north
and two in the south (each separated at $l = 180^{o}$) and examine
nearby ($z \le 0.10$) and distant ($0.10 < z \le 0.14$) clusters
separately. From Figure 2, we see
reasonably fair coverage throughout the entire sky in both redshift
ranges (recall that
the southern right quadrant only goes to $z=0.10$).
Quantitatively, we present in Table 1 the number of clusters available in each
quadrant cataloged by Abell/ACO, and the number
of clusters observed in each quadrant. Note that the fractional
coverages in each of the sections are very similar. The mean fractional coverage
(including both near and far quadrants) is $0.138\pm{0.019}$, so that the
number of clusters within the more distant,
northern right quadrant is only $1.5\sigma$ smaller than the mean.
Table 1 provides clear evidence that the sky coverage for our cluster sample is
not observationally biased towards certain regions.

After accounting for projection effects, line-of-sight anisotropies, X-ray identifications,
HI column density variations, a constant number density, and fair sky coverage,
this is the largest, 
most complete, and fairly sampled distribution of matter in the local
Universe. We assume that clusters are biased tracers of mass (Kaiser 1986; Peacock \& Dodds 1994) and
that in the end, we may compare the shape of our power spectrum to those of typical cosmological models.

\section{Methods and Analyses}
We utilize two different methods to estimate $P(k)$ in redshift-space. Both methods follow
the same basic idea: directly sum the plane wave contributions from each cluster, account for
appropriate weights and the shape of the volume, compute the square of the modulus of each
mode and subtract off the shot noise. The resultant power spectrum is the estimated variance
of the density contrast $|\delta(k)|^2$.
The  power spectrum is accurate only to some limiting scale, specified by $k_{min}$,
which is constrained by the size and shape of the volume examined.
The differences between the two methods arise
when accounting for the weighting scheme and the shape of the volume.
We also point out that Tegmark {\it et al.} (1998) have recently presented
an alternative method for measuring $P(k)$ for large datasets (such as the Sloan
Digital Sky Survey). As discussed in detail in Tegmark {\it et al.}, they
advocate the use of standard Fourier techniques on small scales, a pixelized quadratic
matrix method on large-scales, and also a Karhunen - Loeve (KL) eigenmode analysis to
probe redshift-space anisotropies. While the Tegmark {\it et al.} power spectrum
estimation method is undoubtedly
more refined than the methods used here, we are more interested in comparing results
from the most commonly used techniques (and also allowing our results to be compared
to previous cluster $P(k)$ measurements). Also, the methods described in Tegmark et al.
are designed for large datasets (e.g. several 100,000 points) and we would not expect
a large advantage in our smaller samples.

The first method we use was originally applied by Vogeley et al. (1992),
Park et al. (1994) and da Costa et al. (1994) 
to the CfA2 redshift survey (Geller \& Huchra 1989). This method is also
described in LCRS and Fisher {\it et al.} 1993.
Most recently, this method was used and described by Retzlaff {\it et al.} (1998) on a 
sample of Abell/ACO clusters. 
Briefly, the estimated power spectrum convolved with
the window function can be written as follows:
\begin{equation}
\hat{P}_{a}(k) = {\frac{V}{1-|\hat{W}(k)|^2}}[\hat{\Pi}(k) - \hat{S}].
\end{equation}
The first factor in Equation 1 accounts for the systematic under-estimation of $P(k)$ at
small values of $k$ due to  normalization biases and 
the shape of the window, also known as large-scale power damping (Peacock \& Nicholson 1991). 
The first term in brackets, $\hat{\Pi}$ is the squared-modulus of the Fourier transform of the 
density contrast, $\delta(\bf{r})$, minus the Window Function,
or the estimated power. $\hat{\Pi}$ is a discrete quantity that
includes shot noise $\hat{S}$ which we must subtract off.
The estimate of the power, $\hat{P}_{a}(k)$, is convolved with the
Window function, $\hat{W}(k)$.

In practice, we calculate $\hat{W}(k)$ separately for as many points as is feasible (in this case 
$3\times 10^5$ random points) and average over 1000 directions of $k$. 
The window function  is presented in Figure 3. We calculate $|\hat{W}(k)|^2$ using  points
randomly distributed in our volume and also with the same redshift and extinction
distribution as our real data (the redshift distribution is smoothed with a Gaussian to remove any
large-scale structure). Figure 3 shows that the shape of  $|\hat{W}(k)|^2$ changes
very little as we adjust the random distribution within our volume.
We also calculate $|\hat{W}(k)|^2$ for a volume that
encompasses only one hemisphere out to $z = 0.14$. We see that, as the volume becomes more asymmetric,
significant
differences betweem the window functions appear.
The ``bumps'' seen in $|\hat{W}(k)|^2$ are a direct result
of a volume-limited, spherically symmteric survey, and have little effect on the PS estimation,
so long as their relative heights are much smaller than the largest  $|\hat{W}(k)|^2$ used. 
These ``bumps'' are an indicator that the survey window is spherically symmetric
and that averaging over all directions of ${\bf k}$ is appropriate.
This is not typically the case in previous (non-Abell/ACO)
$P(k)$ analyses (see Tegark (1995) for a good discussion of window functions).

The smallest $k$ that can be accurately probed depends on the value of $|\hat{W}(k)|^2$ and
how it convolves with the real power spectrum (see Lin et al. 1996). Recall, 
\begin{equation}
\langle P_{estimated}(k) \rangle  \propto \int |\hat{W}({\bf k} - {\bf k'})|^2 P_{true}({\bf k'}) k'^2 dk'.
\end{equation}
Ideally, for all values of $k$ probed, the integrand of Equation (2) will be sharply peaked at $k = k'$.
In Figure 4, we plot the integrand of Equation (2) assuming a constant $P(k)$. We find that the
shape of our volume does not affect our analyses for $k >\sim 0.015h$Mpc$^{-1}$.  For $k$ smaller
than this limit, we see that ``leakage'' occurs and power from larger $k$ slips into our measurements.
Figure 4 also shows that if uncorrelated modes of $P(k)$ are required, we should separate
our bins by $\delta k = 0.015$. 
Our choice of $k_{min} =0.015 h$Mpc$^{-1}$ is a conservative
limit, since most past analyses of $P(k)$ have stopped where
$|\hat{W}(k)|^2 = 0.1$ (e.g. Peacock and Nicholson 1991; Vogeley et al. 1992; Retzlaff et al. 1998).
The value of $|\hat{W}(k)|^2$ for our analysis at $k_{min}$ is only $ 0.05h$Mpc$^{-1}$.

The weights for each cluster
originate in the estimation of the density contrast,
\begin {equation}
\hat{\delta}(r)=\frac{1}{N}\sum_i{\frac{\delta^{3}(\bf{r}-\bf{r}_i)}{\phi(r_i)}}  - 1
\end{equation}
where $\phi({\bf r}) = \psi(b)\varphi(z)$ is the selection function which accounts for galactic
obscuration and redshift selection.
We use $\psi(b) = 10^{\gamma(1-csc|b|)}$ with $\gamma = 0.32$ for the latitude selection
function (see Postman, Huchra, and Geller 1992). 
%We use $\varphi(z) = 1$ for  clusters within $z=0.10$ and
%and $\varphi = 0.58$ for clusters with $0.10 < z \le 0.14$.
The selection function in $z$ is determined separately for the three different number density models
used in Figure 1. We find that the choice of number-density fit
has little on the PS estimation.

The second method we use was derived by Feldman, Kaiser, \& Peacock (1994-hereafter FKP).
TED96 use a very similar approach in their analysis of APM clusters.
Here, the
power spectrum is:
\begin{equation}
\hat{P}_b(k) = |F({\bf k})|^2 - P_{shot}
\end{equation}
where $F(k)$ is the Fourier transform of the normalized and weighted galaxy fluctuation field:
\begin{equation}
F(\bf{r})=\frac{w(r)[n_c(r)-{\alpha}n_s(r)]}{[\int d^3r\bar{n}^2(r)w^2(r)]^{1/2}}
\end{equation}
In these equations, $n_c$ and $n_s$ represent the number densities of the cluster sample
and a randomly generated synthetic catalog respectively. 
The number of
points we use in the random catalog is 500 times that of the real data so $\alpha = \frac{1}{500}$
(we note
that there is no difference in the power spectrum results for random catalogs with 100
times as many points). $P_{shot}$ is again, the power due to shot noise from a discrete sample and
is determined as in FKP.
In this method, we model the redshift selection of our random catalogs from
the redshifts of the real data, smoothed with a Gaussian of width $3000$ km s$^{-1}$.
The weights for the individual clusters (real and synthetic) are determined from
\begin{equation}
w_o(r) = \frac{1}{1+n(r)P_{init}(k)}.
\end{equation}
To create the extinction adjusted random catalogs, we draw from regions in
the sky that have the same extinction distribution as our real data using the Schlegel et al. (1998)
maps. We find little difference in our $P(k)$ when we apply no extinction correction.
To determine $n(r)$ used in the weighting factor, we use the three different number density
fits
(as given in Figure 1). Again, we find that the choice of number density fit has little
effect on the PS estimation. 

The weighting scheme for $P_{b}(k)$ depends on {\it a priori} knowledge of $P(k)$ at all
scales. We choose different values of 
$P_{init}(k)$ ( 5, 10, 30, 60$\times 10^4h^{-3}$Mpc$^{3}$)
for the cluster weights and
find that there is little difference in the amplitude  ($\sim 1.5 $ times) 
of $P_{b}(k)$ between
$P_{init} = 5$ and $60\times 10^4 h^{-3}$Mpc$^{3}$ and so we adopt 
$P_{init} = 30\times 10^4 h^{-3}$Mpc$^{3}$ in all further $P_{b}(k)$ results.
We calculate errors on $P_{b}(k)$ using those methods of FKP (equation 2.4.6).

In Figure 5, we compare all of our calculations of $P(k)$.
In the top panel of Figure 5, we plot $P_a(k)$ using the three different number density functions.
We also measure $P(k)$ for the richness adjusted sample.
We plot the same for
$P_{b}(k)$ in the middle panel of Figure 5. In all cases, we find very little difference in our $P(k)$ estimations.
In the bottom panel of Figure 5 we  compare $P_a(k)$  to $P_b(k)$ using the number function as our density fit.
Here, we do see some small differences in the measured power at $k$ less than $0.02h$Mpc$^{-1}$, however
both spectra estimates are within the 1 $\sigma$ error. 
The lack of difference between $P_a(k)$ and $P_b(k)$  
is a direct result of the stability of the methods and the
well defined number density and 
symmetric volume of the cluster sample.

\section{Discussion}

There are two striking results regarding the power spectrum of rich Abell/ACO clusters. (1) While
we do see a dip in power near $k = 0.4h$ Mpc$^{-1}$, it is not statistically significant. 
The measured power spectrum is essentially featureless and there is 
no narrow peak in the power spectrum as has been
reported in E97 and Retzlaff {\it et al.} (1998). (2) The other difference
is that there is increasing power to very large scales ($k = 0.015h$Mpc$^{-1}$ or $\sim 400h^{-1}$Mpc).
In past analyses of the power spectrum, most
authors have reported the (weak) detection of a turnover in the power spectrum (see section 1). However,
the turnover has always occurred very near the largest scales accessible in their volumes. 
[Note: other preliminary analyses of the PS on scales $k < 0.05h$Mpc$^{-1}$ are also showing
this increase in power (Guzzo et al. 1999; Hamilton and Tegmark 2000; Efstathiou and Moody 2000)].
The power spectrum is roughly a power-law on scales $0.02 \le k < 0.10 h$Mpc$^{-1}$ with $P(k) \propto k^{-1.4}$.

In Figure 6, we compare our results to two other cluster sample power spectrum analyses,
the APM cluster sample of TED98, and the $R \ge 0$ Abell/ACO sample of Retzlaff et al. (1998).
Figure 6 shows that the shapes of $P(k)$ for
these three  different cluster samples
are remarkably
similar in the range $ 0.04 \le k \le 0.15 h$Mpc$^{-1}$.
The higher amplitude for our sample of $R \ge 1$ clusters is expected according to
hierarchical clustering schemes (Kaiser 1986) and larger bias found in richer clusters (see
Peacock and Dodds 1994).
We have recalculated the Retzlaff {\it et al.}
(1998) Abell/ACO cluster sample using the methods for $P_a(k)$. We do this in part as a check on
our methods and also to independently confirm their results of a peak near $k = 0.05h$Mpc$^{-1}$
and a turnover thereafter. The Retzlaff {\it et al.} sample includes all Abell/ACO clusters
within $240 h^{-1}$Mpc and outside $|b| \ge 30^{\circ}$. We find 412 clusters which meet this
criteria (compared to their 417 clusters- the difference we attribute to minor variations in
a few cluster redshifts near the survey boundaries). Our results, not surprisingly, are identical
to those published in Retzlaff {\it et al.} (1998) since our method for determining $P_a(k)$ is
identical to theirs. For this determination of $P_a(k)$ ({\it i.e.} using $R =0$ clusters and
a much smaller volume), we also see a peak in the power spectrum at $k = 0.05h$Mpc$^{-1}$
and a turnover thereafter. As pointed out by Retzlaff {\it et al.}, this peak
is not statistically significant. As a further examination of this issue, we plot in Figure 6 
$P_b(k)$ for a smaller cluster sample,
volume-limited in the north and south to $z = 0.10$. For this sample, we can only detect power to
$k_{min} \sim 0.035h$Mpc$^{-1}$. For $k$ greater than $0.035h$Mpc$^{-1}$
we find little difference between this sample and the larger one. But
we can no longer probe on the scales where we expect $P(k)$ to continue its rise.
Thus, one could conclude that a turn-over has been found, when in fact a larger (in size and number) sample
shows that the power continues to rise for $k < 0.03h$Mpc$^{-1}$.

\subsection{Comparisions to Linear Theory}
We also compare our power spectrum results to those of linear theory created by CMBFAST 
(Seljak \& Zaldarriaga 1996). We consider three Cold Dark Matter (CDM) variants, flat, open and
with a vacuum density ($\Lambda$CDM), and a Mixed Dark Matter (MDM) model.
For  the CDM cases, we choose $\Omega_b = 0.02$, in accordance with
Schramm \& Turner (1998). For the open case, we choose
$\Omega_0=\Omega_b + \Omega_{CDM} = 0.2$ in accordance with Bahcall (1997). For the $\Lambda$CDM
model, we choose $\Omega_{CDM} = 0.18$ and $\Omega_{vacuum} = 0.80$ so that
$\Omega_b + \Omega_{CDM} + \Omega_{vacuum} = 1$. For the MDM model,
we choose $H_0 = 50 $km s$^{-1}$Mpc$^{-1}$
with $\Omega_b =0.05$, $\Omega_{CDM}=0.35$ and $\Omega_{\nu}=0.3$
(where $\Omega_{\nu}$ is the massive neutrino density).
The CMBFAST package normalizes
the amplitude of generated spectra to the Bunn and White (1997) four-year COBE normalization. 
However, in this work, we are only concerned with the {\it shape} of the power spectrum.
We are motivated by our assumption that clusters are biased tracers of the mass distribution
and therefore the shape of the cluster power spectrum should be similar to that of the matter
power spectrum.
In Figure 7, we present the amplitude shifted
linear models in comparison to our empirically determined 
power spectra.
As a result of the known similarities in the shapes of the $\Lambda$CDM models and low matter density
open CDM models, we find that both fit the shape of the rich
Abell/ACO cluster power spectrum to $k_{min} = 0.015h$Mpc$^{-1}$ or $400h^{-1}$Mpc extremely
well (see Table 2).
On the largest scales, the 
MDM model lacks power over a wide range of $k$ ($0.015 \le $k$ \le 0.03h$Mpc$^{-1}$) to match
our cluster data.
TED98 found that $\Lambda$CDM linear models did not have
enough power on large scales to match the APM cluster power spectrum. Instead, they find  a much
better fit for
a mixed dark matter (MDM) model.
We
point out that the $\Lambda$CDM model in Figure 7 of TED98 does provide an excellent
fit to the APM cluster data if their last data point at $k = 0.02h$Mpc$^{-1}$ (where
the error is rather large) is excluded.

\section{Conclusion}
The agreement between
the shapes of $P(k)$ for the four different samples shown in Figure 6 (from $k = 0.05$ to $0.15 $h Mpc$^{-1}$),
provides further evidence that clusters are
tracers of the peaks of the underlying luminous 
mass distribution. While there is a
great deal of volume-overlap in these four samples, they are made up of significantly
different luminous objects (from very poor APM clusters to the richest Abell clusters).
For instance, the Retzlaff et al. (1998) Abell/ACO sample
contains at most 253 $R \ge 1$ clusters, while the remaining 218 are $R \ge 0$. 
Our sample contains 637 $R \ge 1$ clusters. The APM sample of 364 clusters, 
contains even fewer $R \ge 1$ Abell clusters ($\sim 40$).
If all groups and clusters
trace the underlying mass distribution in a similar way, then the we would expect their respective
power spectra to be similar in shape, and only the amplitude to vary.
 
Previous analyses of the cluster power spectrum have been plagued
by three major problems: (1) uncertainties in the number density, (2) small volumes,
and (3) irregularly shaped volumes. The sample analyzed in this work greatly improves
upon each of these difficulties. 
Our Abell/ACO sample has a nearly constant number
density throughout the entire volume. This is in stark contrast to most
other sparse tracer surveys (such as the QDOT {\it IRAS} survey power spectrum
of FKP and the Retzlaff {\it et al.} Abell/ACO cluster sample).
Along with the number density, the large size of the volume
and the semi-regular shape of the double-cone geometry, all contribute significantly
to a more accurate determination of $P(k)$ on the largest scales.
The reality of the power on scales $200 - 300h^{-1}$Mpc is also becoming evident
observationally. Batuski {\it et al.} 1999
have recently discovered two filamentary superclusters in the constellation of Aquarius
that are as long as $75h^{-1}$Mpc and $150h^{-1}$Mpc.  As we peer out further into the
local Universe, we continue to find structures on very large scales.

We have presented the redshift-space power spectrum for the largest galaxy cluster sample
compiled to date. This sample has been examined extensively for projection effects, anisotropies,
and observational selection effects and found to be a fair and complete sampling of biased
matter in the local Universe. The volume and shape of the survey provide accurate and robust
measurements of $P(k)$ over the wavenumber range $0.015 \le k \le 0.15 h$Mpc$^{-1}$. From
$k= 0.15$ down to $k = 0.05 h$Mpc$^{-1}$, we
find a similar shape to the power spectrum compared to other cluster samples such as
the APM cluster survey and a smaller sample of $R \ge 0$ Abell/ACO clusters studied by
Retzlaff {\it et al.} (1998). At smaller $k$, we do not find
any statistically significant features in $P(k)$. 
Unlike previous cluster $P(k)$ analyses, we do not find
any strong evidence for a turnover.
We find that $\Lambda$CDM and
low $\Omega_0$ CDM linear models provide excellent fits to the rich cluster power spectrum.

{\bf Acknowledgments} The authors wish to thank Adrian Melott and Daniel Eisenstein
for helpful conversations. We also would like to acknowledge the role of the
the referee, Michael S. Vogeley, for his suggestions on improving the original manuscript. 
We also thank H. Tadros for supplying the APM cluster PS in electronic form.
CM was funded in part by NASA-EPSCoR through the
Maine Science and Technology Foundation.

\begin{deluxetable}{cccccc}
%\tablefontsize{\footnotesize}
\tablenum{1}
\tablewidth{0pt}
\tablecaption{\bf Sky Coverage}
\tablehead{
\colhead{$\ell$ range} & \colhead{ $b$ range} & \colhead{ $z$ range} & 
\colhead{Number (all z)} & \colhead{ Number with} & \colhead{Fraction{\tablenotemark{a}}} \nl
\colhead{ } & \colhead{ } & \colhead{ } &
\colhead{cataloged} & \colhead{observed redshifts} & \colhead{}}
\startdata
$0^o \le \ell < 180^o$ & $30^o \le b \le 90^o$ & $z \le 0.10$ & 636 & 80 & 0.1257 \nl
$0^o \le \ell < 180^o$ & $30^o \le b \le 90^o$ & $0.10 < z \le 0.14$ & 636 & 86 & 0.1352 \nl
$180^o \le \ell < 360^o$ & $30^o \le b \le 90^o$ &$z \le 0.10$ & 503 & 78 & 0.1550 \nl
$180^o \le \ell < 360^o$ & $30^o \le b \le 90^o$ &$0.10 < z \le 0.14$ & 503 & 52 & 0.1034 \nl
$0^o \le \ell < 180^o$ & $-90^o \le b \le -30^o$ &$z \le 0.10$ &608 & 95 & 0.1563 \nl
$0^o \le \ell < 180^o$ & $-90^o \le b \le -30^o$ &$0.10 < z \le 0.14$ & 608 & 84 & 0.1382 \nl
$180^o \le \ell < 360^o$ & $-90^o \le b \le -30^o$ &$z \le 0.10$ & 492 & 75 & 0.1524 \nl
\tablenotetext{a}{Fraction is the Number observed/ Number cataloged.}
\enddata
\end{deluxetable}

\begin{deluxetable}{ccc}
%\tablefontsize{\footnotesize}
\tablenum{2}
\tablewidth{0pt}
\tablecaption{\bf Goodness-of-Fit to Linear Models}
\tablehead{
\colhead{Model} & \colhead{$\chi_{reduced}^2$} & \colhead{DOF}}
\startdata
$\Lambda$CDM ($H_o = 100$km s$^{-1}$) &  0.65 & 8 \nl
Open CDM ($H_o = 100$km s$^{-1}$) & 0.66 & 8 \nl
CDM ($H_o = 100$km s$^{-1}$) & 4.55 & 8 \nl
MDM ($H_o = 50$km s$^{-1}$) & 2.19 &  8 \nl
\enddata
\end{deluxetable}

\begin{figure*}[hl]
\plotone{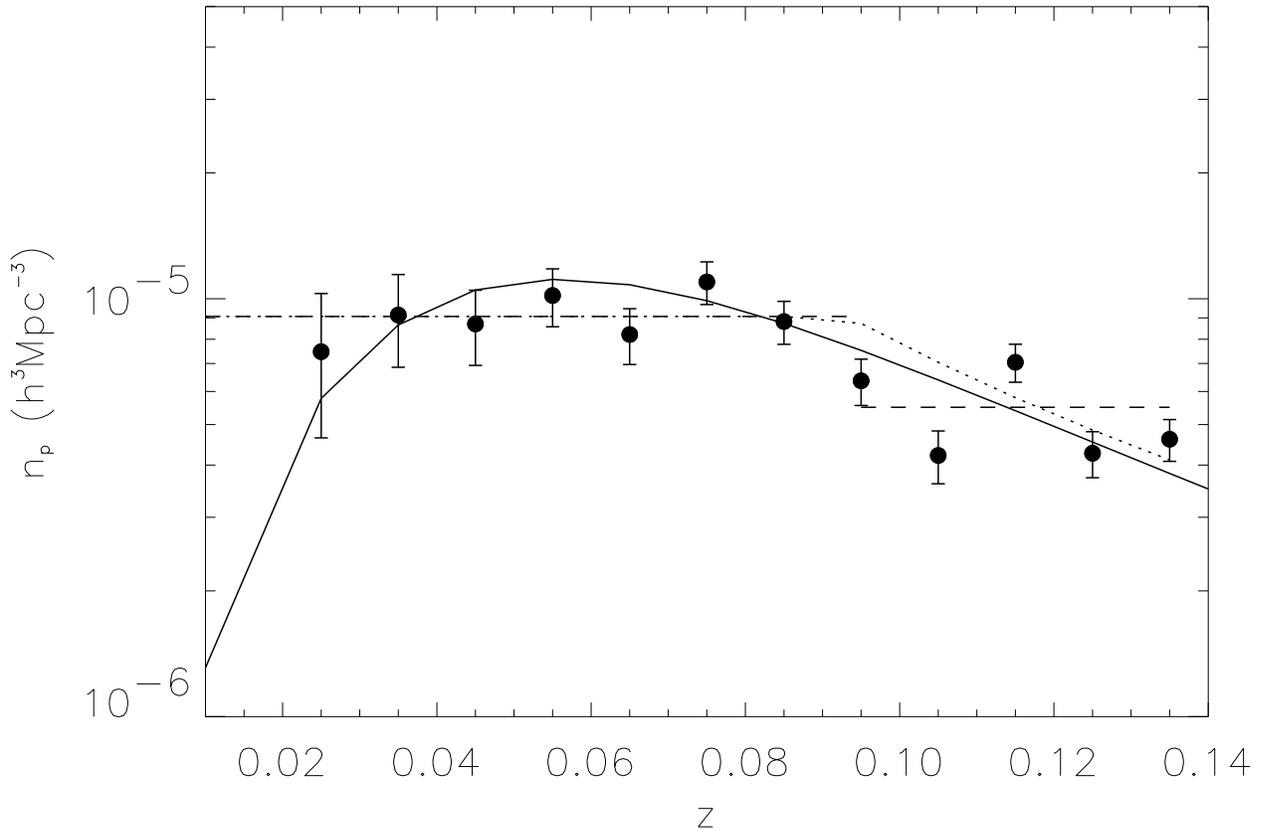}
\caption[]{The proper number density as a function of redshift is presented for the Abell/ACO cluster
sample.  The lines are various fits to the data. The solid-line (having the lowest $\chi^2$) is
a three parameter number function. The dotted-line is for a power-law beyond $z =0.01$. The dashed-line
is for a step function. We find no significant differences in our PS analysis as a function of
the number density function utilized.}
\end{figure*}

\begin{figure*}
\plotone{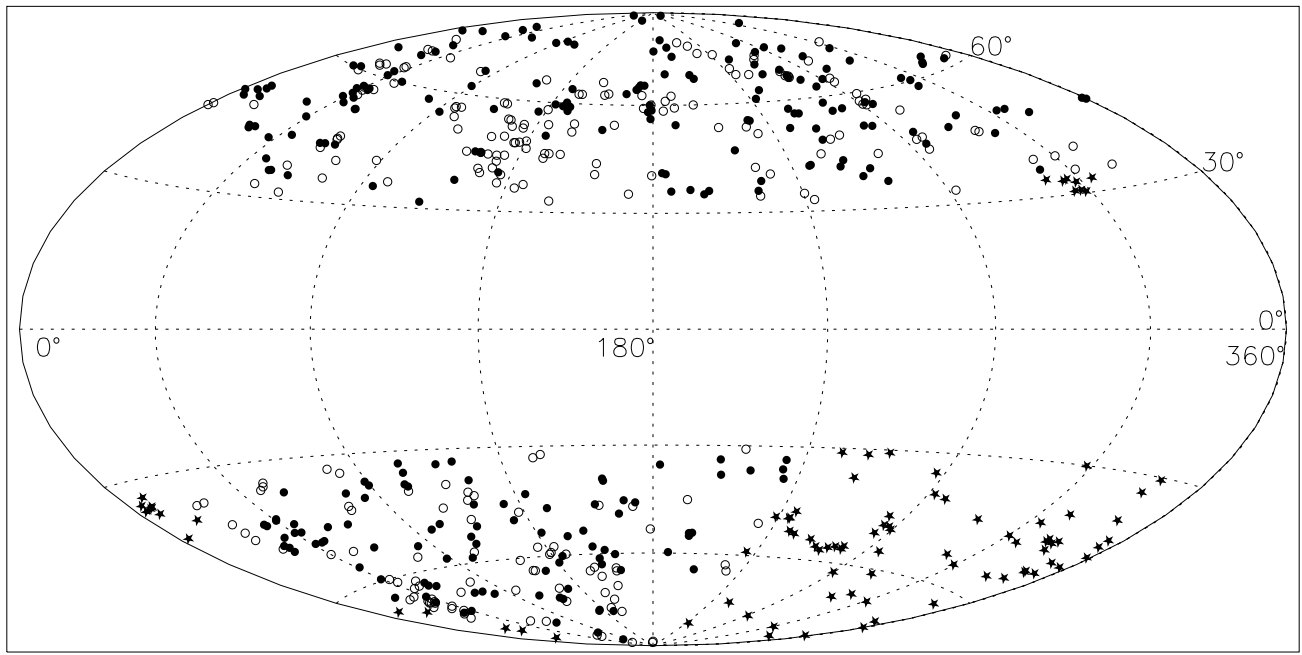}
\caption[]{ A Hammer-Aitoff projection sky-plot of all clusters used in the
power spectrum analysis. Closed circles denote Abell (1958) clusters within $z=0.10$,
open circles denote Abell clusters with $0.10 < z \le 0.14$, and stars indicate
ACO (1989) clusters within $z=0.10$. We have divided our sample into four quadrants
in latitude/longitude and two bins in $z$, to show that the clusters in our sample
have been observed evenly throughout the sky (see Table 1).}
\end{figure*}

\begin{figure*}
\plotone{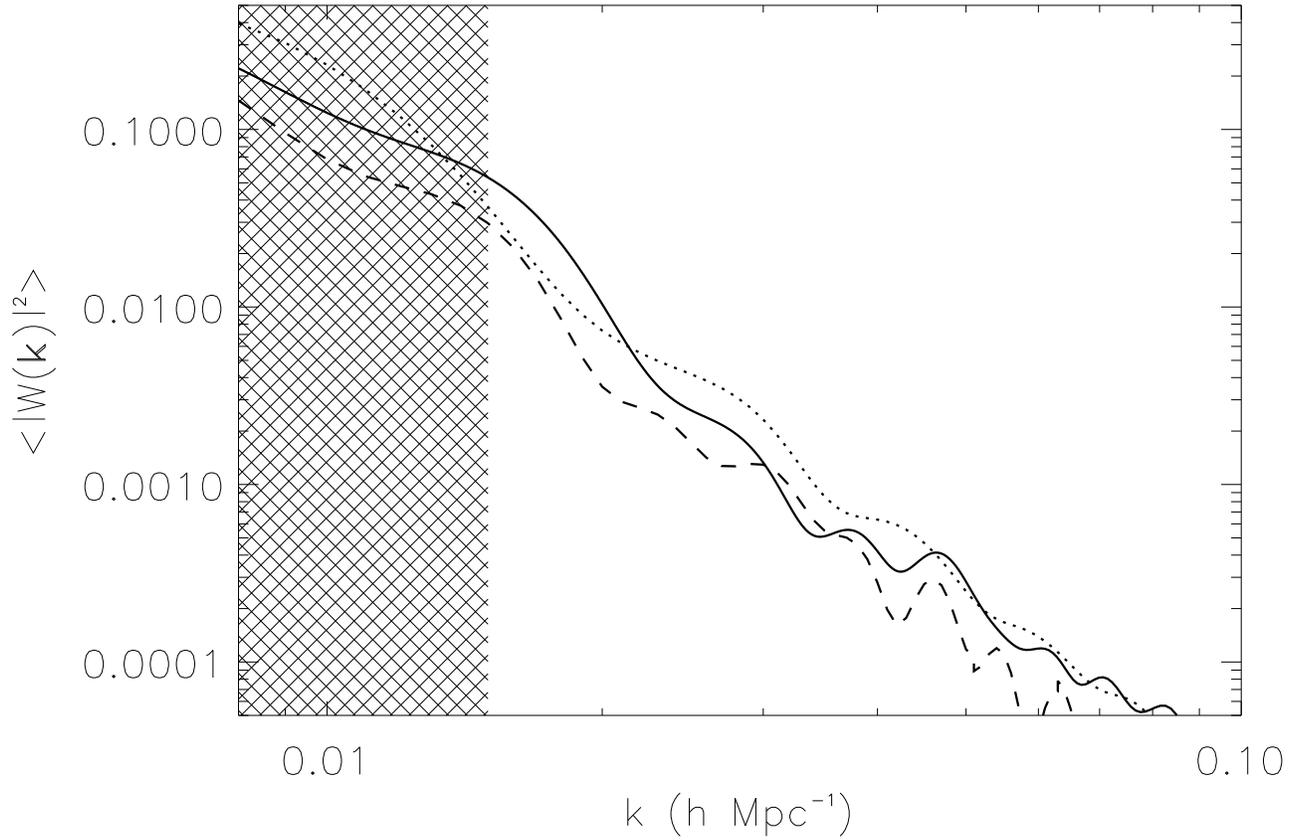}
\caption[]{This is the Fourier window function,  $\langle|\hat{W}({\bf k})|^2 \rangle$ used to calculate $P_a(k)$. We use 300000 points
and 1000 random directions for each $|k|$ to estimate the Fourier transform of the window function. $k$ is in
units of $h$Mpc$^{-1}$. The dashed-line is for a random distribution of points. The solid-line is after
we apply the same redshift and extinction distribution as our real data. The dotted-line is for a highly asymmetric
survey (i.e. one hemisphere to $z = 0.14$).
The hatched region
indicates where our window function prevents an accurate determination of $P(k)$.}
\end{figure*}

\begin{figure*}[hl]
\plotone{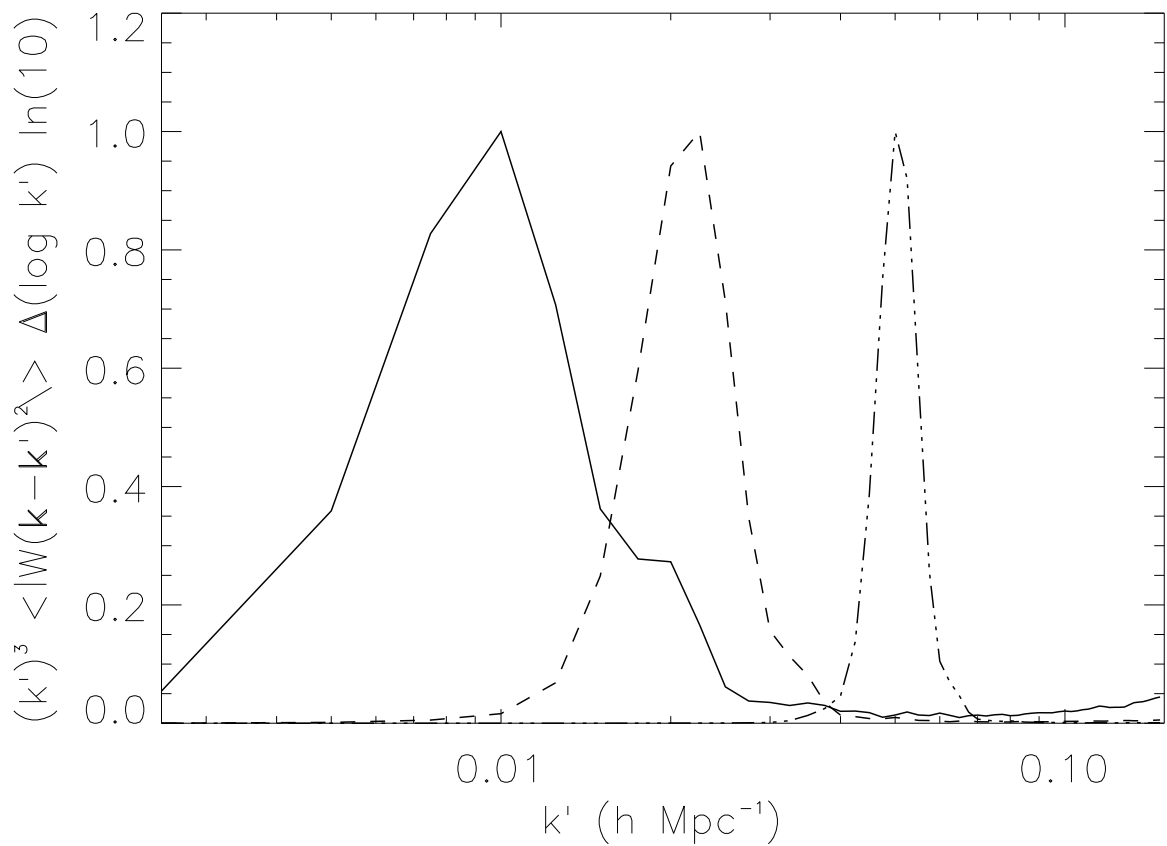}
\caption[]{The integrand of Equation (2) for constant $P(k)$ with arbitrary normalization.
We show three values of $k = 0.01, 0.02, 0.05h$Mpc$^{-1}$. We see that for $ k = 0.01h$Mpc$^{-1}$ there
is ``leakage'' from large $k$. At $k = 0.02h$Mpc$^{-1}$ this leakage is no longer evident, and
so we choose $k_{min} = 0.015h$Mpc$^{-1}$ as the largest-scales we can accurately probe.}
\end{figure*}

\begin{figure*}
\plotone{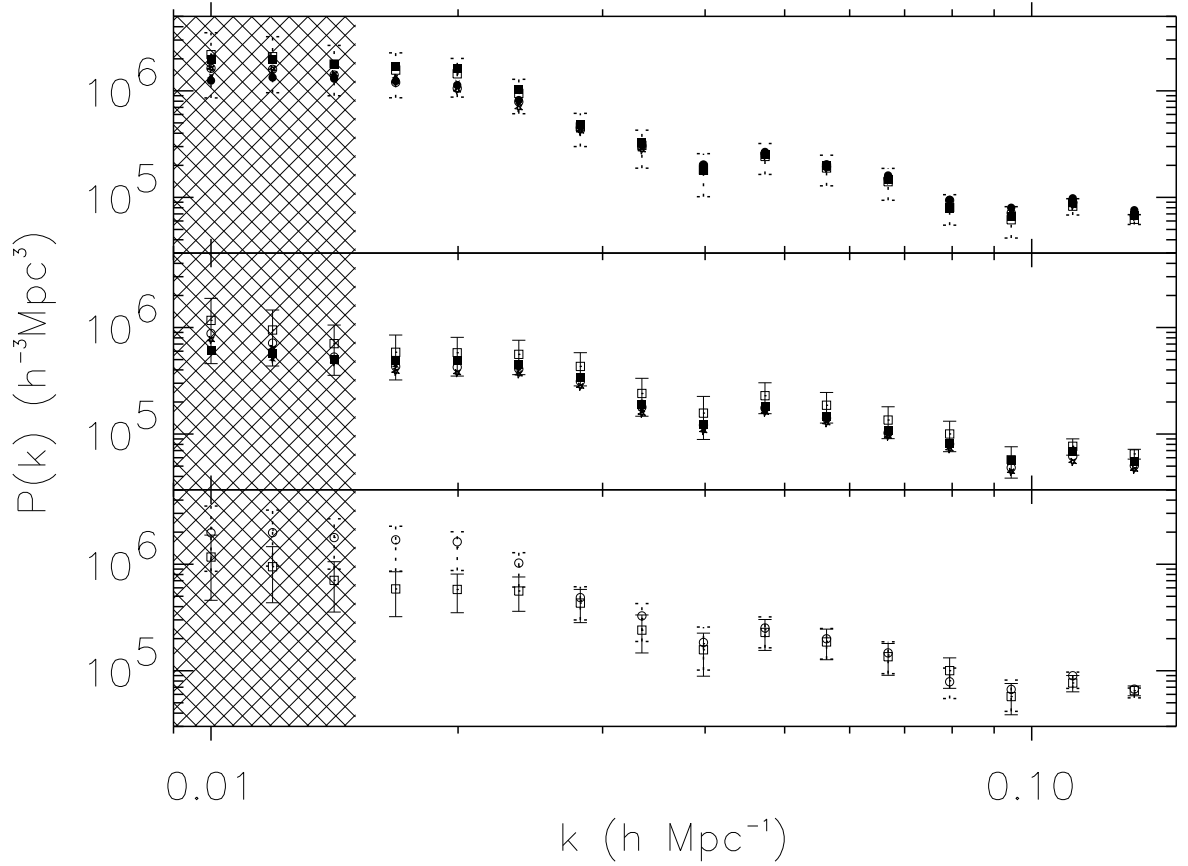}
\caption[]{In the {\bf top} panel we show $P_a(k)$ calculated using the three different number density
functions used in Figure 1. Circles are for the step-function, stars are for the power-law fit,
and squares are for the number function fit. The open circles include all $R \ge 1$ ACO clusters
while the filled symbols are for ACO clusters with $N_{gal} \ge 56$. We use a window function
that models the real data (e.g. in density and extinction). The errors are estimates based
on scaling the errors from $P_b(k)$.
In the {\bf middle} panel we show $P_b(k)$ using the same symbols as the top panel. We use random
catalogs with the same extinction and density distribution as the real data. The error bars
are determined using the FKP method as mentioned in the text.
In the {\bf bottom} panel, we plot $P_a(k)$ (circles) and $P_b(k)$ (squares). The hatched region
indicates where our window function prevents an accurate determination of $P(k)$.
}
\end{figure*}

\begin{figure*}
\plotone{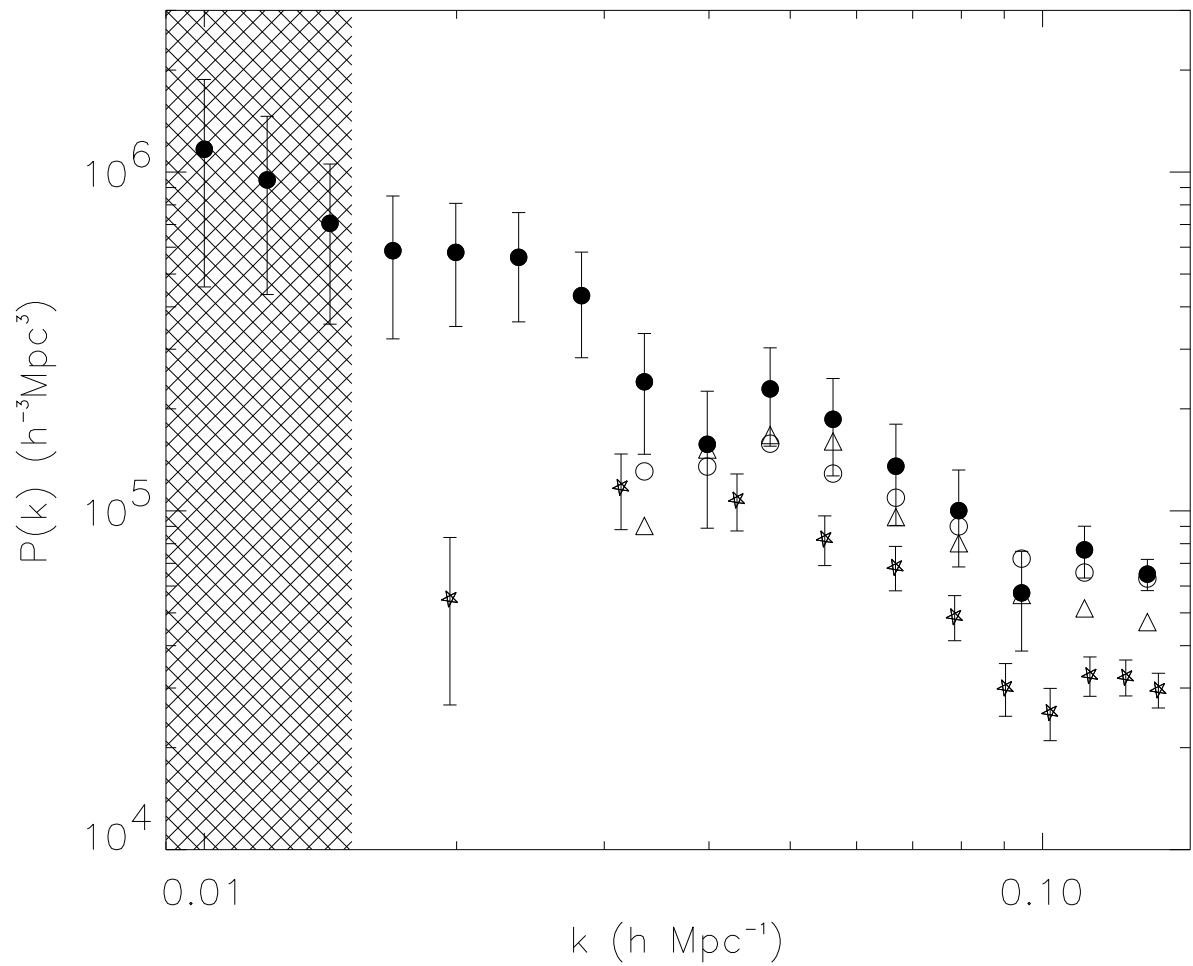}
\caption[]{We compare $P_b(k)$ for Abell/ACO clusters calculated in this work (solid circles) to the smaller
Abell/ACO sample used by Retzlaff {\it et al.} (1998) (triangles) and the APM cluster sample power spectrum
calculated by Tadros {\it et al.} (1998) (stars). The open circles are $R \ge 1$ Abell/ACO clusters within 
$z=0.10$ (north and south). 
}
\end{figure*}

\begin{figure*}
\plotone{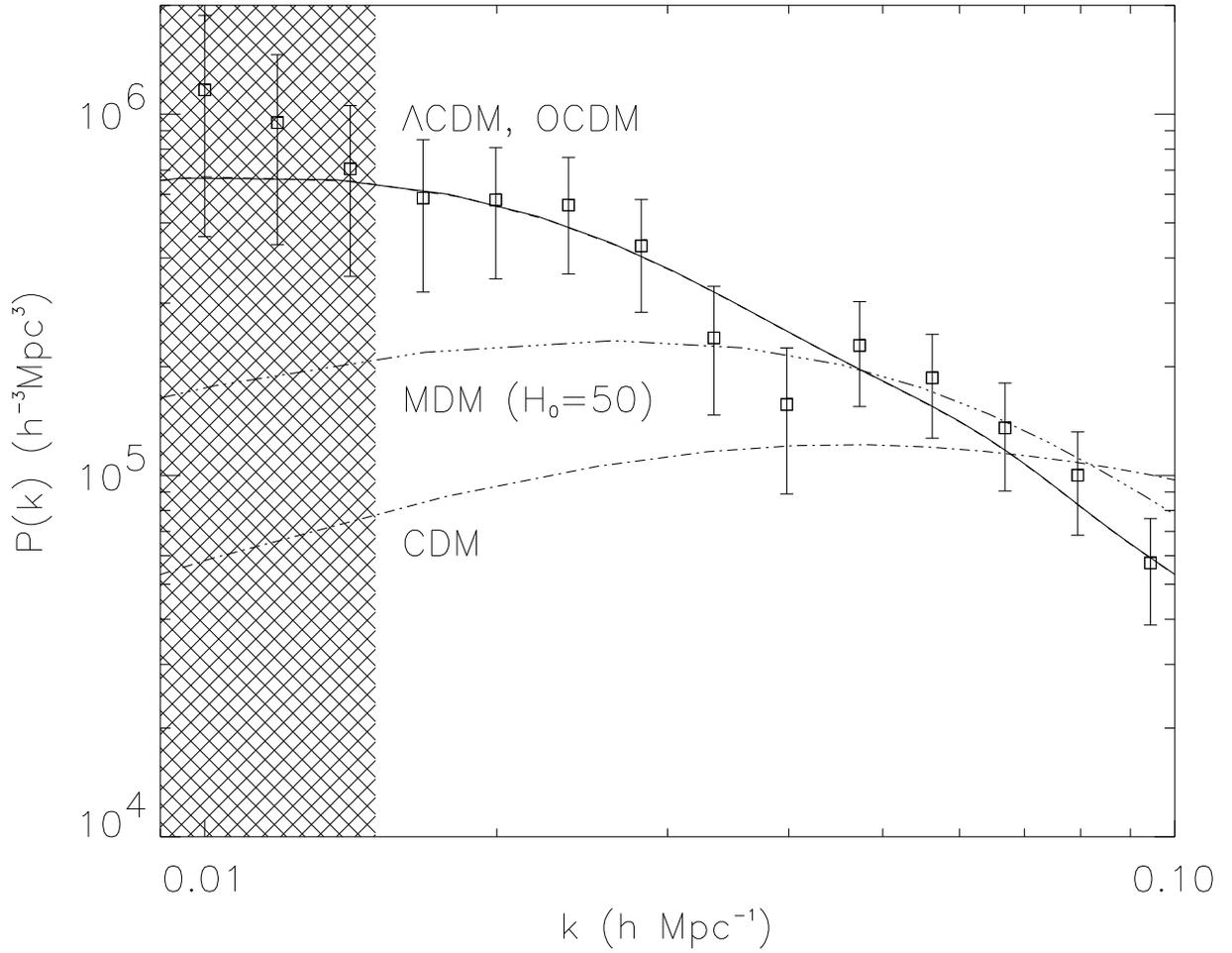}
\caption[]{
We compare $P_b(k)$ to model liner power spectra for a flat CDM 
($\Omega_b=0.02, \Omega_{CDM}=0.98$ with $H_0 = 100 $km s$^{-1}$ Mpc$^{-1}$ {\bf dashed-dot}),
mixed dark matter
($\Omega_b=0.05, \Omega_{CDM} = 0.65$ and $\Omega_{\nu} = 0.3$ with $H_0=50 $km s$^{-1}$ Mpc$^{-1}$
{\bf  dashed-dot-dot}),
open ($\Omega_b=0.02, \Omega_{CDM}=0.18$ with $H_0 = 100 $km s$^{-1}$ Mpc$^{-1}$ {\bf dashed}),
and  lambda 
($\Omega_b=0.02, \Omega_{CDM} = 0.18, \Omega_{vacuum}=0.80$ {\bf solid}) CDM models. }
\end{figure*}

\end{document}